\documentclass{moriond}

\def\be{\begin{equation}}
\def\ee{\end{equation}}
\def\bea{\begin{eqnarray}}
\def\eea{\end{eqnarray}}

\newcommand{\Photo}{\includegraphics[height=35mm]{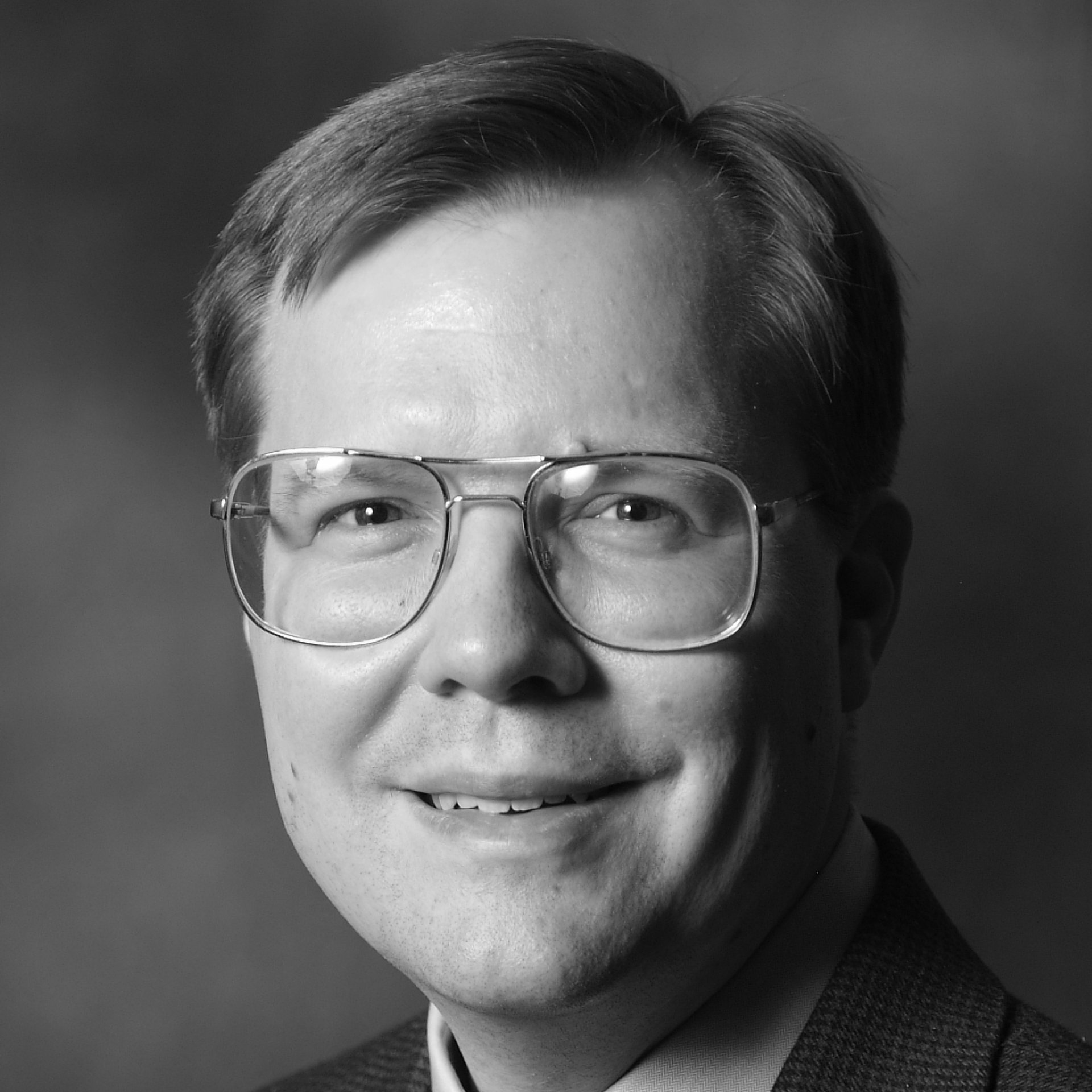}}

\begin{document}
\vspace*{4cm}
\title{BOOSTED-BOTTOM JET TAGGING AND BSM SEARCHES}

\author{ ZACK SULLIVAN }

\address{Department of Physics, Illinois Institute of Technology,\\ Chicago,
Illinois 60616-3793, USA}

\maketitle\abstracts{
  I present a new scheme for tagging boosted heavy flavor jets called
  ``$\mu_x$ tagging'' and its application to TeV-scale physics beyond
  the Standard Model.  Using muons from $B$ hadron decay to define a
  particular combination ``$x$'' of angular information, and jet
  substructure variables, we identify a clean
  ($\epsilon_{\mathrm{fake}}/\epsilon_b\sim 1/100$) good efficiency
  ($\epsilon_b = 14\%$) tag.  I demonstrate the usefulness of this new
  scheme by showing the reach for discovery of 
  leptophobic $Z^\prime\to b\bar b$ and $tH^\pm\to t t b$.}

\section{Introduction}
\label{sec:intro}

\indent\indent As searches for new particles at the CERN Large Hadron
Collider (LHC) shift to TeV-scale energies, observation of their
decays into jets becomes challenging.  Dijet resonances are typically
smaller than the QCD background unless top or bottom tags are applied.
Unfortunately, current $b$-tagging efficiencies degrade (28--15\%)
around 1--2 TeV for light-jet fake rates of 1--2\% \cite{Delphes:3}
(producing low purity $\epsilon_{\mathrm{fake}}/\epsilon_b \sim
1/10$).  At this conference, I presented a new method for flavor
tagging at TeV-scale energies called ``$\mu_x$ boosted-bottom-jet
tagging.'' \cite{Pedersen:2015knf} This method is derived from
kinematic first principles, and provides a $14\%$ efficiency for
$b$-tagging, with a factor of 10 improvement in fake rejection over
existing tags ($\epsilon_{\mathrm{fake}}/\epsilon_b \sim 1/100$).  In
Sec.\ \ref{sec:btag} I summarize the $\mu_x$ definition and cuts, and
plot its transverse momentum- and pseudorapidity-dependent
efficiencies.  In Sec.\ \ref{sec:zp} I briefly describe the reach
provided by $\mu_x$ boosted-$b$ tagging in an analysis for discovery
of a leptophobic $Z^\prime\to b\bar b$.  Following the recent
provocative proposals to measure $H^\pm\to tb$ in $tbH^\pm$-associated
production at the LHC, I use the $\mu_x$ tag in Sec.\ \ref{sec:httb}
to provide a realistic estimate of the reach at a 14 TeV machine.  I
summarize our results in Sec.\ \ref{sec:concl}.


\section{$\mu_x$ boosted-$b$ tag}
\label{sec:btag}

\indent\indent 
In Ref.\ 2 we introduced the $\mu_x$ boosted-$b$
tag, a high purity $b$ tag for use with boosted jets ($p_{T}> 500$
GeV) based on the kinematics of semi-muonic $B$ hadron decay and jet
substructure.  At large momentum, the boost $\gamma_B$ of the $B$
hadron compresses its decay products into a narrow subjet at high
energy.  We define a ``smart'' angular lab frame observable
\begin{equation}
x\equiv \gamma_B \tan\theta_{\mathrm{lab}} ,
\end{equation}
where $\theta_{\mathrm{lab}}$ is the angle between the muon and the
direction of the $B$ hadron in the jet.  For sufficiently boosted $B$
hadrons ($\gamma_B \ge 3$) the lab frame distribution of the muon
count $N$ vs.\ $x$ follows a universal shape.  We define the $\mu_x$
boosted-$b$ tag by first demanding $x < 3$ to capture 90\% of muons
from the $B$ decays.  Then we demand
\begin{equation}
f_{\mathrm{subjet}} \equiv \frac{p_{T \mathrm{subjet}}}{p_{T \mathrm{jet}}} \ge 0.5 ,
\end{equation}
to account for the observation that the hard fragmentation function
for $b$ quarks leads to the $B$ hadron subjet carrying a large
fraction $f_{\mathrm{subjet}}$ of the total jet momentum.

Both cuts in the $\mu_x$ tag depend on identification of the subjet
containing the $B$ hadron.  While exact identification of the $B$ is
not possible, an effective proxy can be found by taking standard
anti-$k_T$ clustered jets with $R=0.4$, and reclustering the muon and
calorimeter towers using a smaller $R = 0.04$.  Following the detailed
reconstruction algorithm of Ref.\ 2, we combine
an identified probable charm subjet remnant, with double the muon
momentum (as a boosted neutrino proxy) to provide the input to
$\gamma_B$ and $p_{T \mathrm{subjet}}$.  A custom $\mu_x$ tagging
module \texttt{MuXboostedBTagging} for DELPHES \cite{Delphes:3} is
available on GitHub.\cite{Delphes:myDelphes}

\begin{figure*}[htb]
\centering
\includegraphics[width=0.4\textwidth,clip]{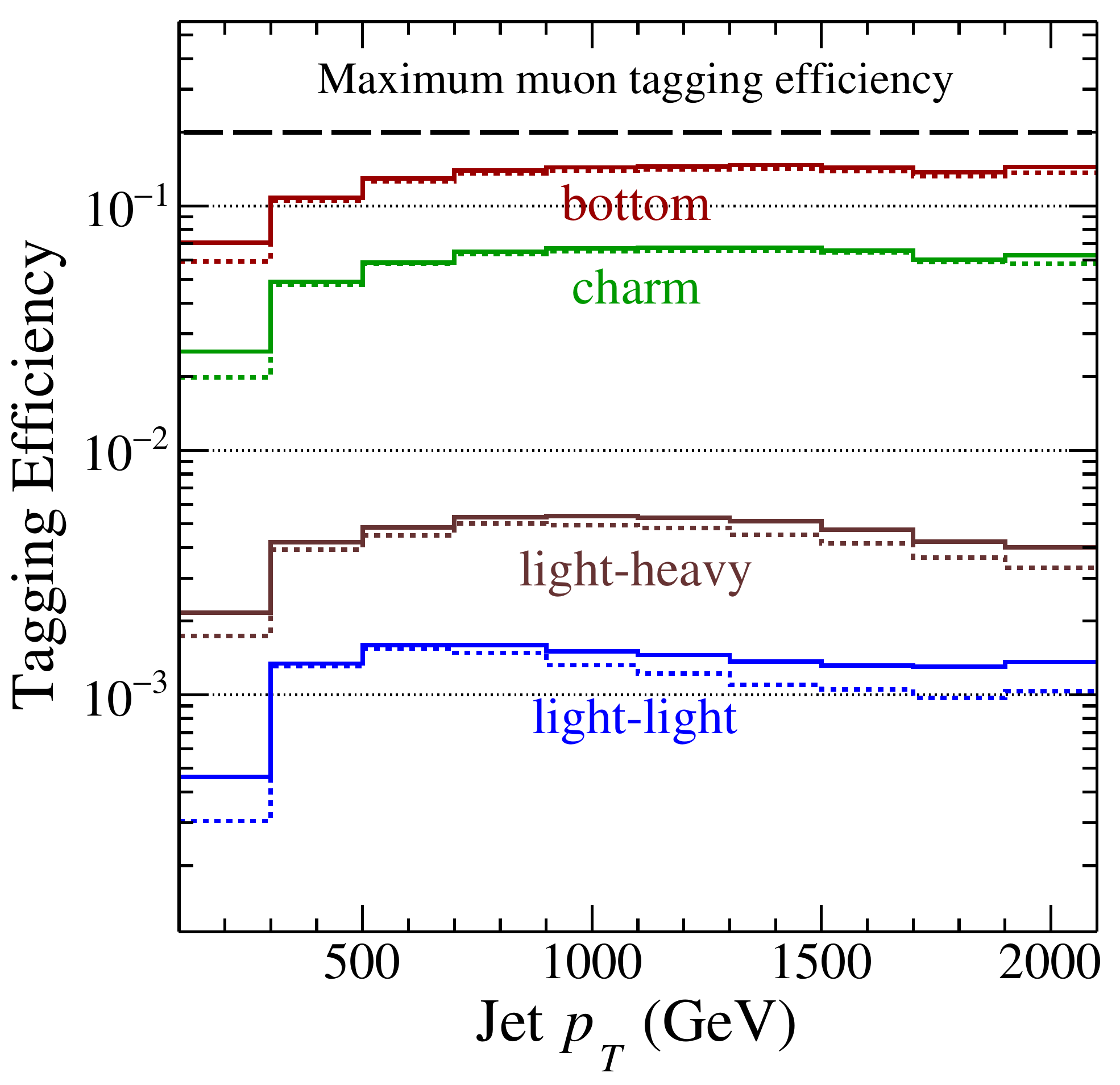}\ 
\includegraphics[width=0.4\textwidth,clip]{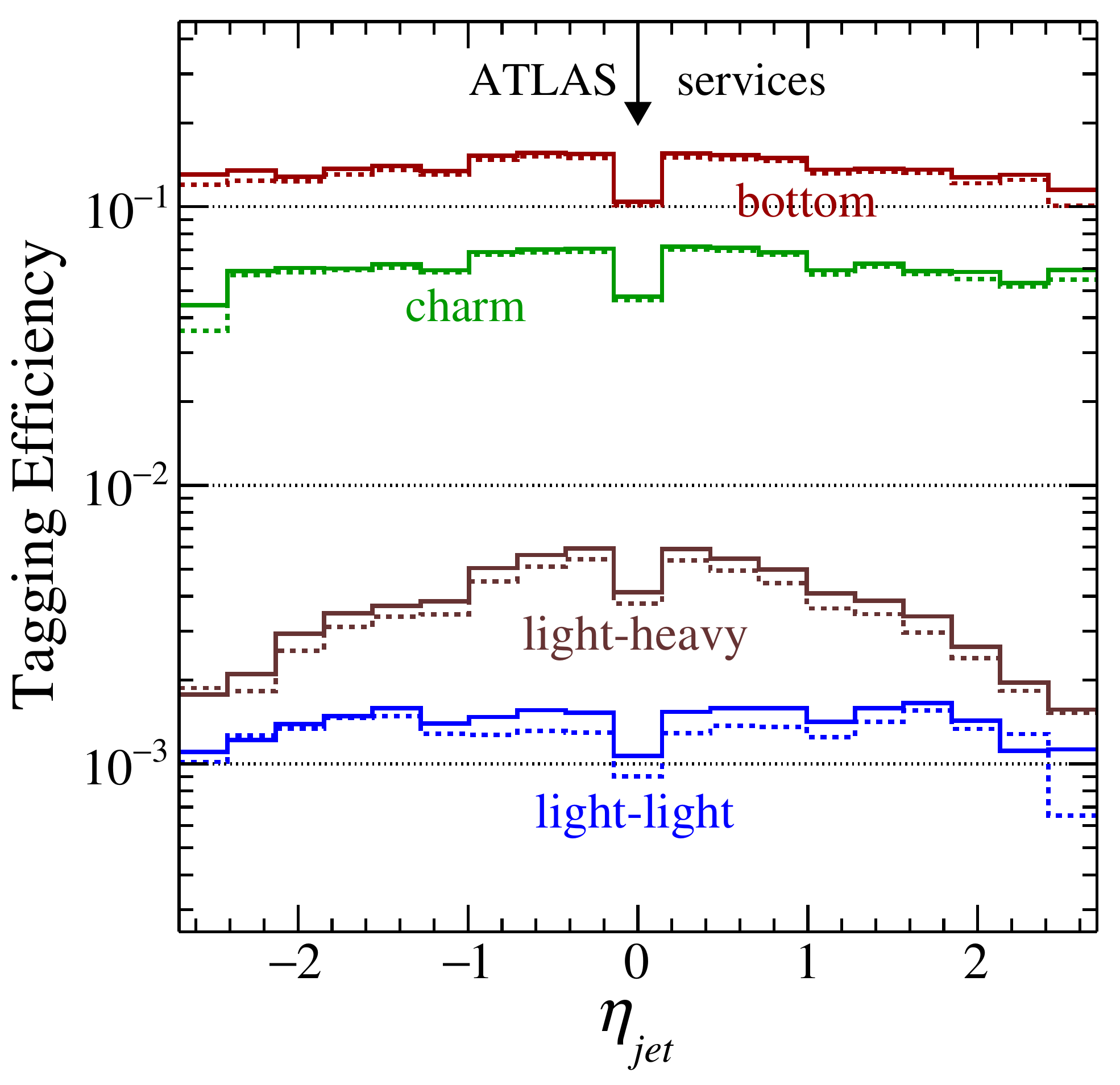}
\caption{$\mu_x$ tagging efficiency vs.\ (left) jet $p_T$ and (right)
  $\eta_{\mathrm{jet}}$.  Solid (dashed) lines include $\mu=0$ (40)
  pileup events.}
\label{fig:eff}
\end{figure*}

In Fig.\ \ref{fig:eff} we see $\mu_x$ tagging efficiencies as a
function of $p_T$ and $\eta$ for bottom jets, charm jets, light-light
jets (where the muon came from a light-flavor hadron), and light-heavy
jets (where a gluon split to $b\bar b/c\bar c$ --- producing
heavy-flavor hadrons in the final state).  The kinematic nature of the
tagging variables leads to nearly flat $p_T$ efficiencies when $p_T >
500$ GeV.  The $\eta$ distribution is also flat except for $B$ hadrons
from gluon splitting.  This leads to the intriguing possibility that
the $g\to b\bar b$ contribution to jets in the Monte Carlo could be
calibrated using the rapidity dependence of these highly-boosted jets.

\section{Leptophobic $Z^\prime\to b\bar b$ at the LHC}
\label{sec:zp}

\indent\indent
New neutral vectors, generically called $Z^\prime$ bosons, appear in
many BSM models.  In cases where the decay to leptons is suppressed,
we look to tag heavy flavors to overcome the QCD dijet background.  We
examine the reach \cite{Pedersen:2015knf} at a 13 TeV LHC for a
leptophobic $Z^\prime$ decaying to $b\bar b$ or $c\bar c$ using a
$U(1)^\prime_B$ Lagrange density \cite{Dobrescu:2013coa}
\begin{equation}
{\cal L}=\frac{g_B}{6}Z_{B\mu}^{\prime}\bar{q}\gamma^{\mu}q ,
\end{equation}
with flavor-independent coupling to quarks.

The signal and backgrounds are simulated using a MLM-matched MadEvent
sample \cite{MadGraph:5} and CT14llo PDFs \cite{CTEQ:CT14} fed through
PYTHIA 8 \cite{Pythia:8.1} into DELPHES.  We demand one or two $\mu_x$
tags, $|\eta_j| < 2.7$, and $\Delta\eta_{jj} < 1.5$.  We reconstruct a
dijet mass out of the two leading-$p_T$ jets, and look for a resonance
in the mass window $[0.85,1.25]\times M_{Z^\prime_B}$.

\begin{figure*}[htb]
\centering
 \includegraphics[width=0.8\columnwidth,clip]{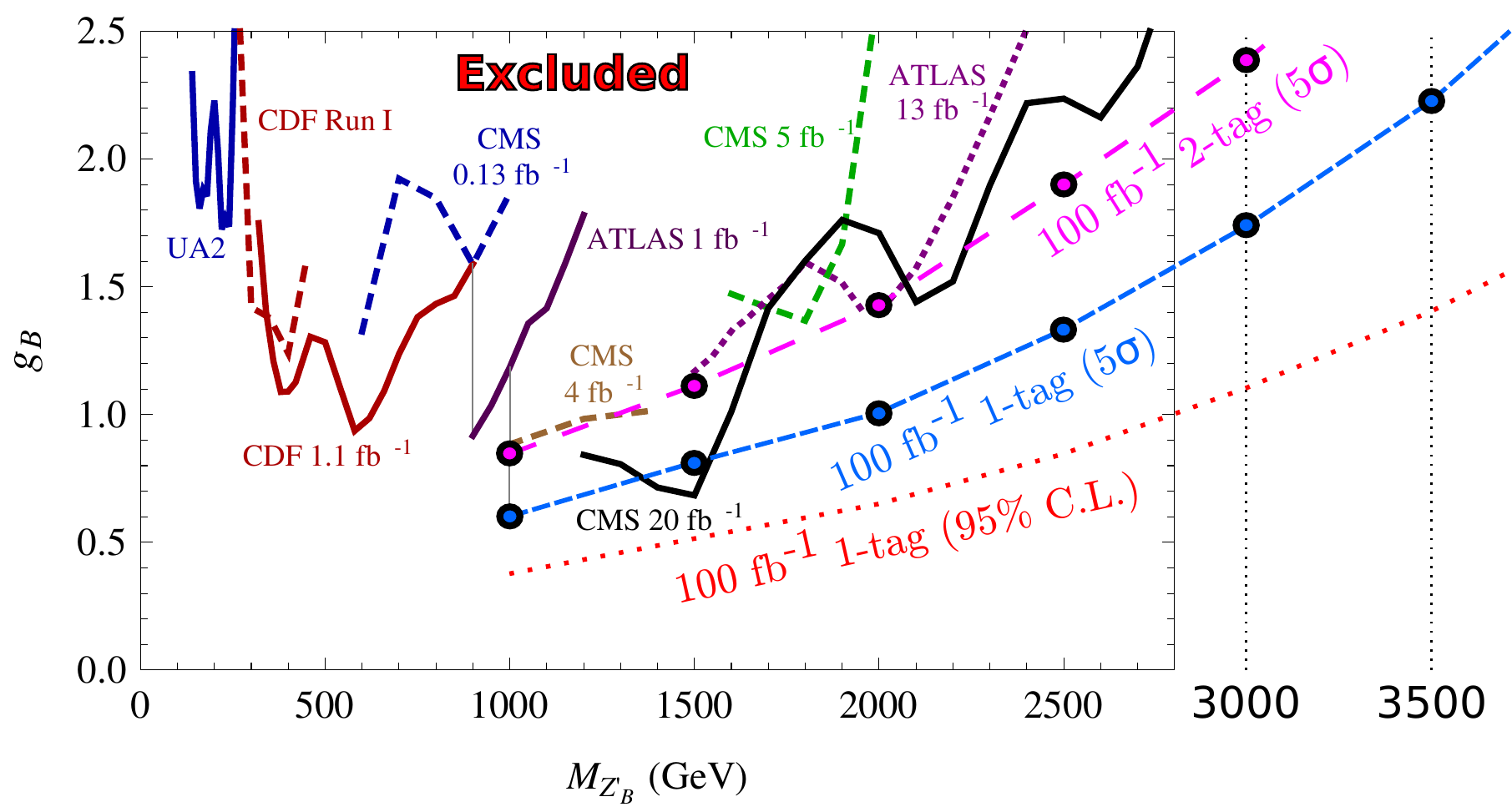}
 \caption{$5\sigma$ discovery reach for a leptophobic $Z^\prime$ with
   universal coupling and one or two boosted-$b$ tags at a 13~TeV LHC
   compared to exclusion limits from Ref.\ 4.  Also shown is the 95\%
   C.L.\ exclusion reach of the one-tag analysis.}
\label{fig:zplimits}
\end{figure*}

We see the reach for $5\sigma$ discovery of this leptophobic
$Z^\prime$ in Fig.\ \ref{fig:zplimits} for a two-tag, and one-tag
inclusive sample, \cite{ismd15} compared to current exclusion limits
from Ref.\ 4.  In 100 fb$^{-1}$ of integrated
luminosity at 13 TeV, a two $b$-tag analysis could discover a
$Z^\prime$ of 3~TeV if the universal coupling $g_B \sim 2.5$.  The
single-tag inclusive search is even more effective --- allowing for
discovery up to 1~TeV above mass limits from Run I.  In the absence of
a discovery, the one-tag search would set a 95\% C.L.\ exclusion that
can access $g_B$ couplings a factor of 2 smaller than current limits,
and masses up to 2~TeV higher.

\section{Associated top--charged Higgs production $tH^\pm\to ttb$}
\label{sec:httb}

\indent\indent
In the MSSM, associated production of charged Higgs with a top quark
produces a final state rich in $b$ jets.  Recent excitement was
generated by a claim \cite{Hajer:2015gka} that the ``wedge region'' in
$\tan\beta$ ($\tan\beta\sim 6$ where the $h^0$ shares equal coupling
to top and bottom) could be explored up to 2~TeV in $H^\pm$ mass at a
14 TeV LHC through the channel $tbH^\pm\to tb(tb)$.  In contrast,
others~\cite{Craig:2015jba} found that even 500 GeV could not
be probed.  We explore \cite{hplus} whether the $\sim 2$ TeV limit can
be reached in the wedge region, and how $\mu_x$ tagging performs in
this final state.

\begin{figure*}[htb]
\centering
\includegraphics[width=0.4\textwidth,clip]{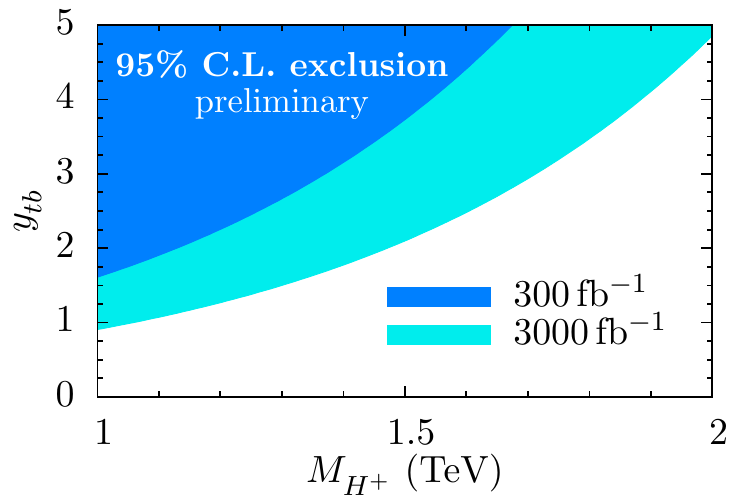}\ 
\includegraphics[width=0.4\textwidth,clip]{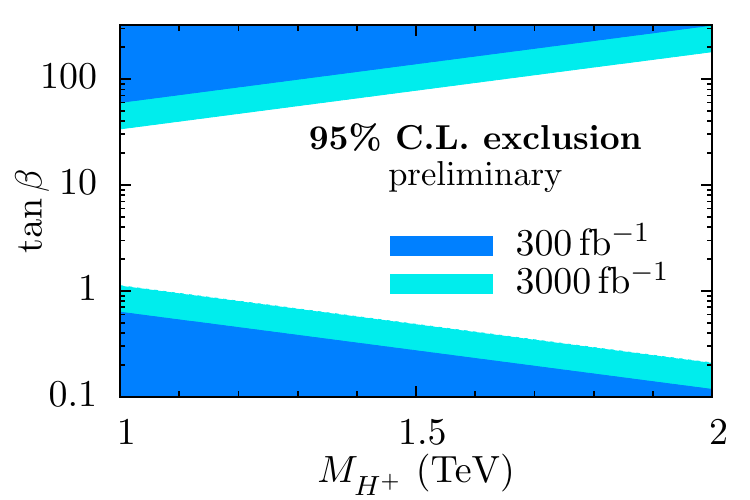}
\caption{95\% C.L.\ exclusion limits that can be reached as a function
  of $H^\pm$ mass at a 14 TeV LHC for (left) $H^\pm tb$ Yukawa coupling
  $y_{tb}$ and (right) $\tan\beta$.}
\label{fig:httb}
\end{figure*}

Using MLM-matched samples for $tH^\pm \to ttb$ generated in MadEvent,
showered in PYTHIA, and reconstructed in DELPHES, we look for final
states involving one boosted-top tag, one $\mu_x$ boosted-$b$ tag, and
a fully reconstructed $t\to bl\nu$ decay (with a normal low-energy $b$
tag).  The background is dominated by fake tags from $t\bar tj$ and $t
jj$ that we take as measured from CMS data.  After cuts on the relative
$p_T$ and angle of the two leading jets, we find $S/B\sim 1/10$.  A
preliminary estimate of the reach in $H^\pm tb$ Yukawa coupling
$y_{tb}$ and $\tan\beta$ are shown vs.\ $H^\pm$ mass in Fig.\
\ref{fig:httb}.  Our analysis appears to extend the results of Ref.\
10 up to 2~TeV --- meaning the wedge remains.  It
appears the wedge region will need to wait until a 100~TeV machine.

\section{Conclusions}
\label{sec:concl}

\indent\indent
At Moriond 2016 I presented the new $\mu_x$ boosted-bottom jet tag and
its applications to some important searches for new physics at the
LHC.  Combining angular information $x$ from $B$ hadron decay with jet
substructure $f_{\mathrm{subjet}}$ in TeV-scale jets allows for clean
extraction of signals for $Z^\prime$ and MSSM charged Higgs above
backgrounds.  We find that the reach for leptophobic $Z^\prime$
discovery at a 13~TeV LHC is about 1~TeV higher than current limits.
If a $Z^\prime$ is not found, 95\% C.L.\ exclusion limits can be set
up to 2~TeV higher, or $g_B$ couplings a factor of 2 smaller, than
the current limits.  

Despite recent excitement, the search for MSSM charged Higgs in the
mid-$\tan\beta$ ``wedge'' region in $tH^\pm \to t(tb)$ will remain
elusive.  The signal appears to be too small when realistic tagging
efficiencies are applied.  A 100~TeV collider is likely needed to fill
this region.  On the other hand, the $\mu_x$ tag could be used to
immediately improve the existing searches for $W^\prime\to t\bar b$ in
the boosted-top and boosted-bottom channel. \cite{Duffty:2013aba}  The
$\mu_x$ boosted-bottom jet tag is a powerful new tool in the
exploration for physics beyond the Standard Model.

\bigskip

\section*{Acknowledgments}
This work was supported by the U.S.\ Department of Energy under award
No.\ DE-SC0008347.

\end{document}